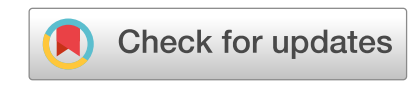

OPEN

# Forecasting consumer confidence through semantic network analysis of online news

Andrea Fronzetti Colladon[1]✉, Francesca Grippa[2], Barbara Guardabascio[3], Gabriele Costante[1] & Francesco Ravazzolo[4,5]

**This research studies the impact of online news on social and economic consumer perceptions through semantic network analysis. Using over 1.8 million online articles on Italian media covering four years, we calculate the semantic importance of specific economic-related keywords to see if words appearing in the articles could anticipate consumers' judgments about the economic situation and the Consumer Confidence Index. We use an innovative approach to analyze big textual data, combining methods and tools of text mining and social network analysis. Results show a strong predictive power for the judgments about the current households and national situation. Our indicator offers a complementary approach to estimating consumer confidence, lessening the limitations of traditional survey-based methods.**

Monthly reports of the nation's level of consumer confidence can offer a first understanding of the consumers' sentiment and predict their spending. Consumer confidence has been traditionally associated with objective political and economic conditions and external factors like news media coverage. Indeed, previous research has shown that survey data, such as the Consumer Confidence Index (CCI), can successfully support the forecasting of economic variables released with a substantial delay e.g.,[1–3]. Survey data are often used as initial conditions for macroeconomic models or for modifying a baseline distribution to match certain moment conditions of interest given by the survey e.g.,[4–6]. Cascaldi-Garcia and colleagues[7] also showed that opinion surveys, such as the CCI, are particularly important for nowcasting economic variables released with a substantial delay and push forward the idea of predicting economic opinions surveys.

However, despite the general attention given to consumer confidence surveys, their reliability in providing information about the future path of household spending is still not entirely explored[8]. It has been demonstrated how the predictive power of consumer confidence surveys is influenced by factors including economic conditions, current and past political situations, trust in the government, and the influence of the news industry[9–11]. The media, both mainstream and digital, can influence how consumers feel about the economy. Barsky and Sims[12] found that the relationship between consumer confidence and consequent activity is almost entirely reflective of the news component. In a series of studies on the role of media in influencing the stock market and the financial performance of companies, Tetlock[13,14] used a Bag-of-Words approach to quantify the language used in financial news stories and found that, contrary to popular belief, media pessimism weakly predicts increases in market volatility. In another study, Tetlock, Saar-Tsechansky and Macskassy[15] found that negative language used in financial press articles can predict low earnings for firms, suggesting that the words used in news stories are not superfluous information, but rather, they capture essential aspects of a company's fundamentals that are otherwise difficult to quantify. Li's study[16] on the usage of the terms "risk" and "uncertain" in a company's annual reports highlights the importance of paying attention to the language used in financial reports. By analyzing the words chosen by companies, investors can gain insight into the level of risk associated with the company's operations. Market prediction and consumer behavior mechanisms that rely on online text mining are only now beginning to be thoroughly investigated, thanks to the significant advancements in computational processing power and network speed in recent years[17]. In our study, we adopt a Big Data methodology to forecast consumer confidence, looking at the role of online news and its influence on consumer confidence. We investigate how online news—as reported in digital newspapers and other online sources—influences consumer confidence using a methodology that relies on an indicator that calculates the importance of economic-related keywords (ERKs) appearing on digital news media. This is a departure from the time-consuming manual content analysis of economic news used in the past[18]. Our focus is on the Italian Consumer Confidence Climate index, which provides an indication of the optimism and pessimism of consumers who evaluate the Italian general economic situation and report their expectations for the future. We chose an indicator of semantic importance, called Semantic Brand Score (SBS), which calculates the relative importance of one or more keywords in the news[19]. We selected this indicator because of its ability to forecast various outcomes, from financial market trends[20] to election results[21] and tourism demand[22]. Based on methodologies drawn from social network analysis and text mining, the semantic importance of keywords is calculated in terms of their prevalence, i.e., frequency of word occurrences; connectivity, i.e., degree of centrality of a word in the discourse; and diversity, i.e., richness and distinctiveness of textual associations. The approach we use in our study is different from past research that focused on the evaluation of news sentiment e.g.,[23,24]. We have implemented a new integrated semantic index as a measure of semantic significance. This metric has been proven to be more informative than sentiment analysis, which can be subject to variable error rates and reliability issues[25], and represents a valuable tool for analyzing and understanding relationships among words in a corpus[20,22].

[1]Department of Engineering, University of Perugia, Via G. Duranti 93, 06125 Perugia, Italy. [2]Northeastern University, 360 Huntington Avenue, Boston, MA 02115, USA. [3]Department of Economics, University of Perugia, Via Alessandro Pascoli 20, 06123 Perugia, Italy. [4]Faculty of Economics and Management, Free University of Bozen-Bolzano, Universitätsplatz 1 - Piazza Università 1, 39100 Bozen-Bolzano, Italy. [5]Department of Data, BI Norwegian Business School, Nydalsveien 37, 0484 Oslo, Norway. ✉email: andrea.fronzetticolladon@unipg.it

This study contributes to the discussion on online media's role in shaping consumer confidence. By providing an alternative method based on semantic network analysis, we investigate the antecedents of consumer confidence in terms of current and future economic expectations. Our approach is not intended to replace the information obtained from traditional tools but rather to supplement them. For instance, we may use consumer surveys in conjunction with our methods to gain a more comprehensive understanding of the market.

Section "The connection between news and consumer confidence" delves into the impact of news on consumers' perceptions of the economy. Section "Research design" outlines the methodology and research design employed in our study. Section "Results" showcases the primary findings, subsequently analyzed in Section "Discussion and conclusions".

## The connection between news and consumer confidence

Effective news coverage plays a crucial role in shaping the current and future expectations of individuals. Both digital and mainstream media provide information that can significantly impact people's economic evaluations of present and future conditions and influence economic decisions. The information disseminated through news channels can significantly impact the way people perceive the economy, leading to changes in their spending habits, investment decisions, and overall economic behavior[26,27]. The news may influence consumer confidence, especially when people are exposed to ambiguous messages[11], or when media coverage does not fully reflect economic conditions, or when it is biased by partisanship[9,28]. For example, Damstra and Boukes[29] investigated the impact of the real economy on economic news in Dutch newspapers and confirmed that the description of economic reality offered by the media is skewed to the negative, which in turn affects people's economic expectations about the future, but not their current evaluations. Other studies show the role of rumors in shaping consumer response and spending[30], while others demonstrate how the tone of economic news may influence consumer confidence, with a slight difference between prospective versus retrospective economic evaluations[18,31]. For example, Boukes et al.[18] found that consumers' retrospective evaluations were not influenced by the tone of the news. Other studies explored the effect of the negativity bias on consumer confidence and demonstrated how consumers react only to bad news[10]. The negativity bias, well documented in social psychology, political science, and economics[32,33], is at the basis of this asymmetry in response to bad versus good news: negative information often has a more profound effect on the formation of impressions than positive information. As a result, negative information can have a lasting impact on our perceptions and judgments. Other scholars have challenged the negativity bias and the asymmetric response of consumers. In a study examining the relationship between media reporting of economic news and consumer confidence in the United States, Casey and Owen[31] found evidence of positive and negative consumer confidence asymmetries.

Empirical studies have demonstrated how alternative methods based on textual analysis are more reliable and could complement and reduce the limitations of survey-based methods to describe current economic conditions and better predict a household's future economic activity. For example, a recent study conducted on the accuracy of Swiss opinion surveys revealed that the level of survey bias varies significantly depending on the policy areas being measured. The study found that the strongest biases were observed in areas related to immigration, the environment, and specific types of regulation.

This information is crucial for policymakers and researchers who rely on public opinion surveys to inform their decisions. By understanding the potential biases in survey results, they can make more informed decisions and develop more effective policies[34]. Song and Shin[35] have recently conducted a study on sentiment analysis of South Korean news articles using a lexicon approach. Their findings have demonstrated the potential of news as a valuable source for developing alternative economic indicators that can supplement traditional Consumer Confidence indices. News data is not only cheaper to acquire: its advantages, compared to monthly national surveys, include the ability to observe consumer trends at a more granular level, with more data points, and the ability to capture the social and economic impact of specific issues through a broader perspective[36]. Additional empirical evidence confirms the complicated relationship between consumers and news reported by the media. Through an investigation of the association between consumer spending for durable goods and consumer confidence, Ahmed and Cassou[37] found that news has a relevant impact on confidence during economic expansions, though it is generally not important during economic recessions.

Contributing to this stream of research, we use a novel indicator of semantic importance to evaluate the possible impact of news on consumers' confidence.

## Research design

**Consumer confidence index survey and selection of keywords.** Consumer confidence climate is a monthly economic indicator that measures the degree of optimism perceived by consumers regarding the overall state of the economy and their financial situation, evaluated through their saving and spending habits. Its value is high when consumers spend more and save less and low when consumers save more and spend less. Its trend typically increases when the economy expands and decreases when the economy contracts, reflecting the outlook of consumers with respect to their ability to find and retain good jobs according to their perception of the current state of the economy and their financial situation.

In Italy, the Consumer Confidence Climate survey is composed of a set of questions designed to assess consumers' perceived optimism or pessimism around the Italian economic situation and their expectations for the future. Survey participants provide their opinion about future unemployment, current and future households' financial situation, current and future possibility of savings, current opportunities for durable goods purchases, and current family budget. The answers to nine questions are aggregated, and the result is reported in a seasonally adjusted index[38]. The Consumer Confidence Climate can be broken down into four sub-indices released by the Italian Institute of Statistics (ISTAT). These indices are: the Economic Climate, the Personal Climate, the Current Climate, and the Future Climate. The *Economic Climate* Index considers consumers' current assessment and future expectations regarding the general economic situation in Italy, as well as their outlook on future unemployment. The index of *Personal Climate* takes into account various factors that impact a household's financial well-being. These include the current financial situation, savings, significant purchases of durable goods, and the family budget. The *Current Climate* index analyzes various factors that impact the Italian economy, including the current financial situation of households, their savings, expenditures on durable goods, and family budget. Finally, the *Future Climate* includes questions related to the foreseen future of the Italian general economic situation, the households' financial situation, unemployment expectations, and savings. We downloaded the target series data from the Italian National Institute of Statistics (ISTAT) website (https://www.istat.it).

From the Consumer Confidence Climate survey, we extracted economic keywords that were recurring in the survey's questions. We then extended this list by adding other relevant keywords that matched the economic literature and the independent assessment of three economics experts. The inclusion of external experts to validate the selection of keywords is aligned with the methodology used in similar studies[39]. These keywords, translated from Italian, include home, rent, income, pensions, savings, credit, loans, interest rates, prices, market, job, competition, economy, public sector, politics, institutions, basic necessities, global, family, trust, discomfort/distress, consumer, education degree, purchase, car, PC, and holidays. These keywords provide insight into the concerns and priorities of Italian society. From the basic necessities of home and rent to the complexities of the economy and politics, these words refer to some of the challenges and opportunities individuals and institutions face. We also considered their synonyms and, drawing from past research[20,40], we considered additional sets of keywords related to the economy or the Covid emergency, including singletons—i.e., individual words—such as Covid and lockdown.

Table 1 shows the full list of ERKs, with the *RelFreq* column indicating the ratio of the number of times they appear in the text to the total number of news articles.

| ERK | Relative frequency | ERK | Relative frequency | ERK | Relative frequency |
|---|---|---|---|---|---|
| Covid | 7.902% | House | 14.031% | Retirement pension | 2.601% |
| Lockdown | 1.472% | Purchase | 8.637% | Saving | 2.052% |
| Unemployment | 2.176% | Rent | 0.920% | Vacation | 2.045% |
| Quantitative easing | 0.080% | Car | 6.025% | Competition | 1.503% |
| Economic crisis | 1.093% | Incomes | 2.935% | Spread | 0.735% |
| Interest rate | 3.550% | Employee loans | 0.011% | Rating | 0.443% |
| Monetary politics | 0.249% | Loan | 2.919% | Euro-group | 0.147% |
| European community | 1.469% | Family | 9.263% | Coronabond | 0.011% |
| Financial markets | 9.537% | Trust | 1.949% | Eurobond | 0.033% |
| EU | 8.192% | Discomfort | 0.689% | Sure | 0.087% |
| Unions | 6.431% | Job | 40.451% | European investment bank | 1.117% |
| Strikes | 1.182% | PC | 1.066% | Oil | 1.203% |
| Deficit | 1.151% | Politics | 19.345% | Gold | 1.622% |
| Taxation | 1.725% | Public sector | 0.069% | Troika | 0.055% |
| Btp-bot | 0.478% | Economy | 17.787% | Euro | 24.193% |
| Prices | 2.123% | Education degree | 2.110% | Italian public retirement system | 0.881% |
| Italian stock exchange | 1.048% | Consumers | 3.483% | GDP | 1.943% |
| Bank of Italy | 0.324% | Needs | 0.395% | Confindustria (National Industrial Association) | 1.265% |
| smartworking | 0.087% | Global | 2.818% | | |
| Junkbond | 0.002% | Institutions | 4.071% | | |

**Table 1.** ERK frequencies.

Computational methods have been recognized as unable to understand human communication and language in all its richness and complexity[41]. Aligned with contemporary approaches to semantic analysis[39,42], we have integrated computational methods with traditional techniques to analyze online text. Our methodology incorporates algorithmic measures to systematically gather news data.

Telpress International B.V.—a company that collects online news from multiple web sources, including mainstream media sites and blogs—provided access to online news data. The final sample comprised over 1,808,000 news articles published between January 2, 2017, and August 30, 2020. Our textual analysis focused solely on the initial 30% of each news article, including the title and lead. This decision aligns with previous research[21] and is based on the understanding that online news readers tend only to skim the beginning of an article, paying particular attention to the title and opening paragraphs[43,44]. As a robustness check, we ran our models on the full text of the articles but found no significant improvement in results. This information is also useful because it enables faster data analysis.

**A new index of importance for economic keywords.** The Semantic Brand Score (SBS) is a composite indicator measuring semantic importance, which combines text mining and social network analysis methods. It is applied to (big) textual data to evaluate the importance of one or more 'brands' or, more in general, words or groups of keywords[19]. Its analytical power extends beyond commercial brands. A brand may refer to commercial products, personal brands, a company's core values, or concepts related to societal trends[21]. The SBS indicator is composed of three dimensions: prevalence, diversity, and connectivity. This index builds upon the relationships among words in any given text. Its first component, prevalence, measures how frequently an economic-related keyword is used in the online discourse. The more a word appears in online news, the more readers will remember and recognize it, which could ultimately influence their opinions and behaviors. For instance, consider the scenario where the phrase "economic crisis" is repeatedly featured in the media. This could potentially instill a sense of fear and uncertainty among the general public, leading them to believe that their employment status or financial stability is in jeopardy. However, the importance of a keyword does not only depend on its frequency of occurrence but also on its association with other keywords in the text. For instance, the utilization of the term "economic crisis" in a broad context and its association with various other words can significantly impact people's emotions and actions. Conversely, if the term is solely linked to a job crisis or a war occurring in a far-off land, it can also trigger a shift in people's attitudes and behaviors. To account for these scenarios, the SBS indicator includes other dimensions, diversity and connectivity, which help evaluate how heterogeneous and strong are the associations to an ERK and how much that concept can bridge connections among other terms/concepts in the discourse. The concept is that the greater the frequency of a keyword within a discourse, and the more it is enriched with associations, the more it will be retained and have a significant impact.

To calculate diversity and connectivity, we analyzed the semantic networks generated from online texts using the SBS BI web application[45], and we relied on the computing resources of the ENEA/CRESCO infrastructure[46]. The first step of the computational process was to apply common text pre-processing routines[47]—such as tokenization, removal of stop-words, and removal of word affixes, known as stemming[48]. The second step was to build a social network of co-occurring words for each week of news and study them through social network analysis[49]. Figure 1 illustrates an example of output visualized using the following sentence attributed to Adam Smith (The Theory of Moral Sentiments, 1759): "*The same principle, the same love of system, the same regard to the beauty of order, of art and contrivance, frequently serves to recommend those institutions which tend to promote the public*

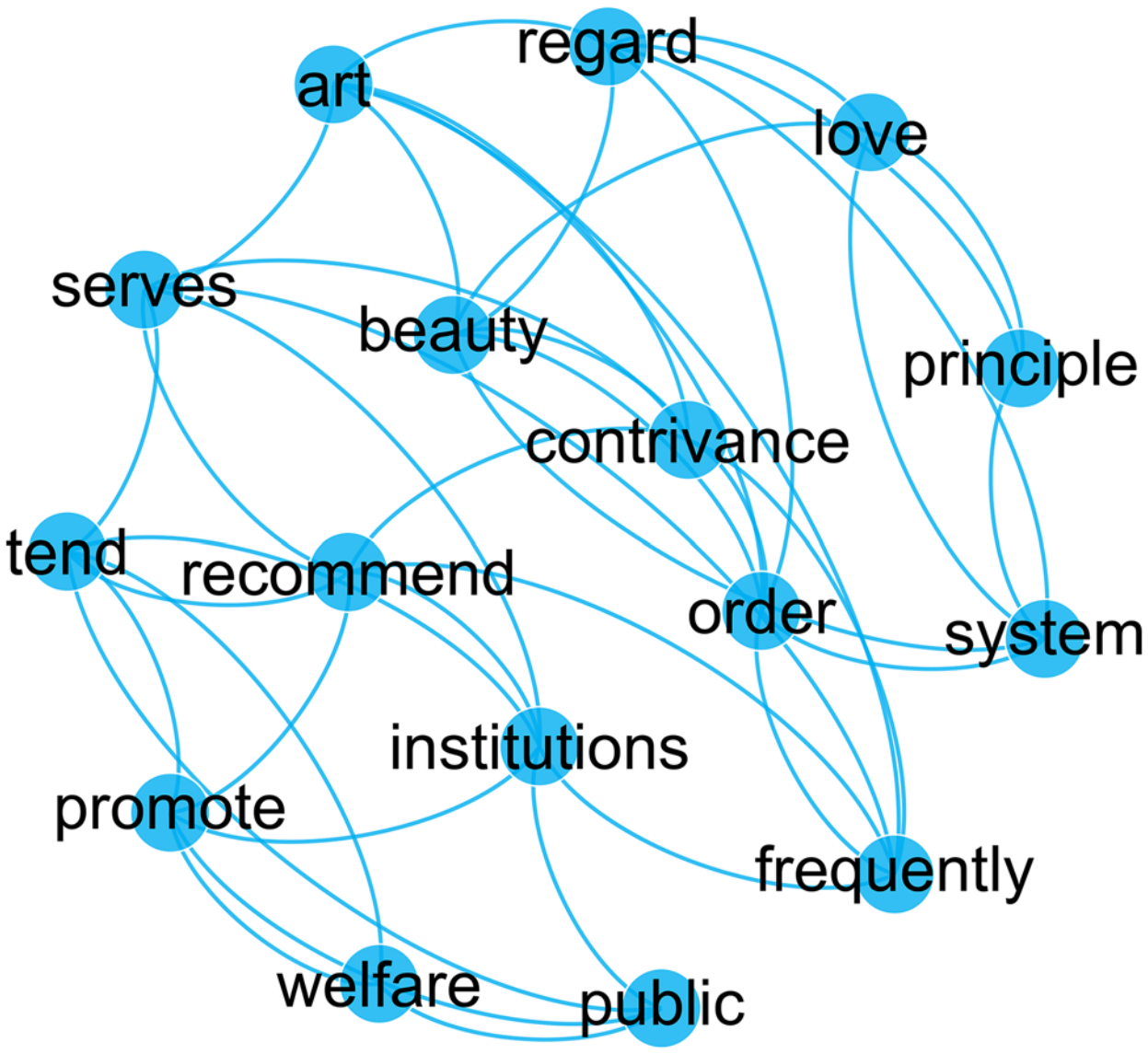

**Figure 1.** Example of semantic network analysis.

*welfare*". In order to improve readability, we labeled nodes before stemming, and we set the co-occurrence threshold to maximum three words.

Diversity is a dimension of the SBS index that considers the relationship of economic keywords with the other words in the text. This is related to the construct of brand image[50] and to the idea that, when associations are less common and in a high number, the keyword is more important[19,51]. We operationalized diversity through the following formula, based on the metric of distinctiveness centrality[52]:

$$Diversity\ (i) = \sum_{\substack{j=1 \\ j \neq i}}^{n} \log_{10} \frac{(n-1)}{g_j} I(w_{ij} > 0).$$

In general, we consider a graph G, made of $n$ nodes (words) and $E$ edges (word links), associated with a set of connection weights $W$. In the formula, $g_j$ is the degree of node $j$, which is one of the neighbors of node $i$ (the one for which diversity is calculated). $I(w_{ij} > 0)$ is an indicator function that is equal to 1 when the edge connecting nodes $i$ and $j$ exists, i.e., when $w_{ij} > 0$, and is equal to 0 when this edge is missing.

The last dimension of the SBS, connectivity, is measured as the weighted betweenness centrality of the ERKs[53,54] and represents their 'brokerage power', i.e. how much each keyword can serve as a bridge to connect other terms and topics in the discourse[19]. The connectivity formula is based on the analysis of the shortest paths connecting each pair of nodes[49]:

$$Connectivity\ (i) = \sum_{j<k} \frac{d_{jk}(i)}{d_{jk}}$$

where $d_{jk}$ is the number of shortest network paths connecting nodes $j$ and $k$ (calculated using edge weights) and $d_{jk}(i)$ is the number of those paths that include node $i$.

The final SBS indicator was calculated by summing the standardized scores of its components, considering all the words in the corpus for each timeframe. Aligned with past studies, e.g.[19,21], we used an equal weighting scheme and carried out standardization by subtracting the mean and dividing by the standard deviation, as in the following formula:

$$SBS\ (i) = \frac{PR_i - \overline{PR}}{std(PR)} + \frac{DI_i - \overline{DI}}{std(DI)} + \frac{CO_i - \overline{CO}}{std(CO)}$$

where PR is prevalence, DI is diversity, and CO is connectivity. We also tested different approaches, such as subtracting the median and dividing by the interquartile range, which did not yield better results.

Lastly, we calculated the language *sentiment* of all articles as a control variable and a possible additional predictor of the Consumer Confidence Index and its dimensions. Sentiment was computed using the SBS BI web app[45], which uses a lexicon similar to VADER[55] for the Italian language. Sentiment scores range from $-1$ to $+1$, with $-1$ indicating very negative article content and $+1$ the opposite.

**Granger causal relationships between keywords and consumer confidence.** The Consumer Confidence series have a monthly frequency, whereas our predictor variables are weekly data series. In order to use the leading information coming from ERKs, we transformed the monthly time series into weekly data points using a temporal disaggregation approach[56]. The primary objective of temporal disaggregation is to obtain high-frequency estimates under the restriction of the low-frequency data, which exhibit long-term movements of the series. Given that the Consumer Confidence surveys are conducted within the initial 15 days of each month, we conducted a temporal disaggregation to ensure that the initial values of the weekly series were in line with the monthly series. To obtain weekly values, we applied a cubic spline interpolation[57–59]. Figure 2 illustrates the disaggregated series we obtained.

To measure whether the SBS indicators offered relevant information to anticipate our economic variables, we performed Granger Causality tests. In general, a time series is said to Granger-cause another time series if the former has incremental predictive power on the latter. Therefore, Granger causality provides an indication of whether one event or variable occurs prior to another. We also looked at the cross-correlation of the target series with our predictors (i.e., ERKs series) to see if they were in phase (positive signs of cross-correlation) or out of phase (negative sign)[60,61].

Figure 3 outlines the methodology employed in our research design. We started by identifying the Economic Related Keywords (singletons or word sets). We then calculated the SBS indicators to measure the keyword's importance and applied Granger causality methods to predict the consumer confidence indicators.

## Results

In this section, we discuss the signs of cross-correlation and the results of the Granger causality tests used to identify the indicators that could anticipate the consumer confidence components (see Table 2). In line with past research, e.g.[62,63], we dynamically selected the number of lags using the Bayesian Information Criteria. The models indicate that 61% of the semantic importance series of ERKs Granger-cause the *Personal* component of the Consumer Climate index, while only 34% Granger-cause the *Future* component and 27% the Current component. It is not surprising that average consumers have a better understanding of their personal situation when responding to questions but may be less informed about economic cycles. When answering questions about their own financial situation, individuals are likely to have a more accurate understanding of their personal circumstances. However, when it comes to broader economic trends and cycles, the average consumer may not have the same level of knowledge or expertise. This is understandable, as economic cycles can be complex and difficult to understand without specialized training or experience. Interestingly, this representation of the current situation comes from online news, which may report what is currently happening more than depicting future scenarios—which may directly impact consumers' opinions and economic decisions.

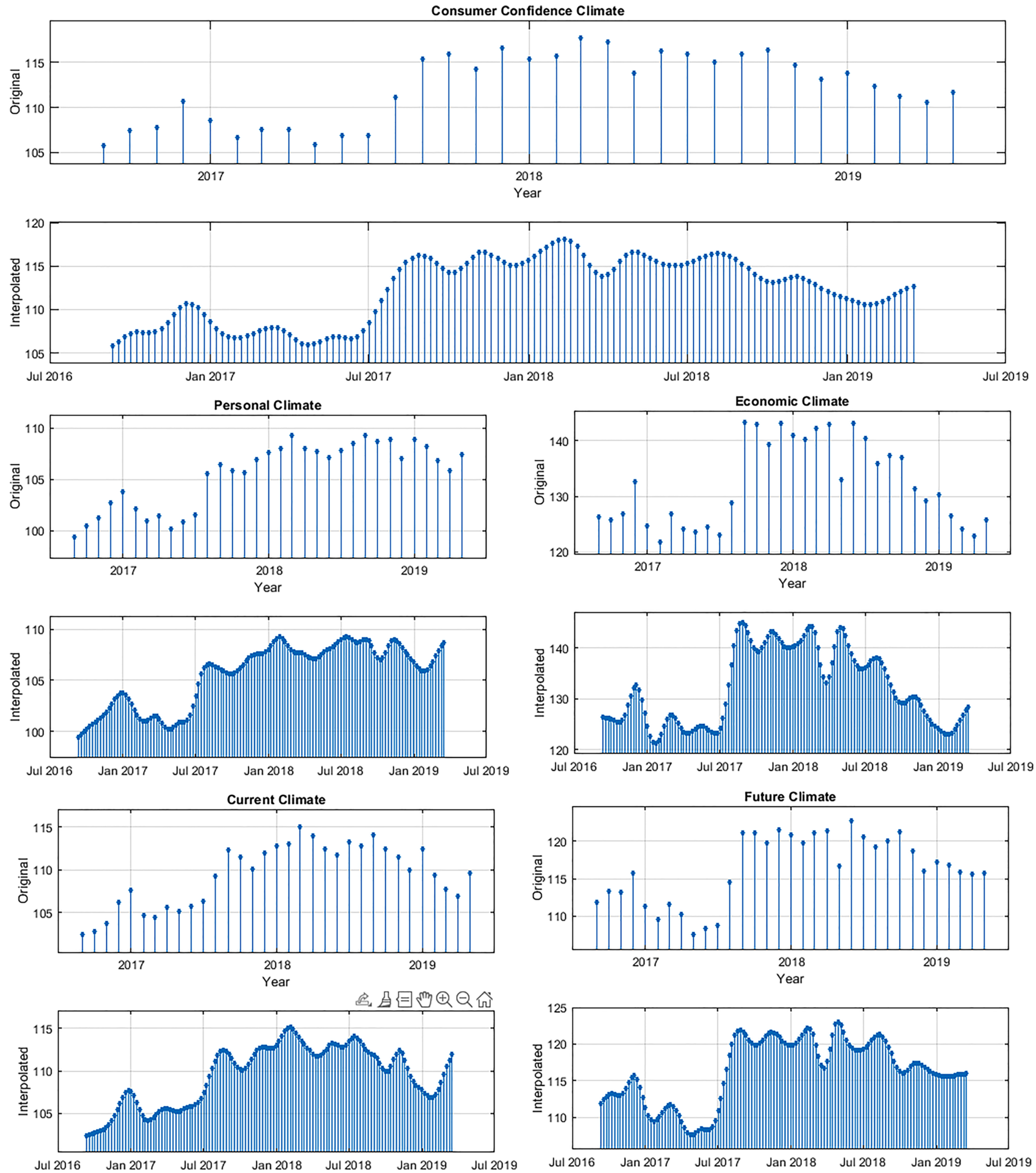

**Figure 2.** Temporal disaggregation of consumer confidence series.

Among the most significant concepts strongly associated with the consumers' confidence in the future, we find keywords such as educational degree, purchasing ability, and a list of European programs to support Italy during the recession (e.g., *Troika*, *Sure*).

**Figure 3.** Research design framework.

Among the political and institutional keywords primarily associated with a perceived deterioration of consumers' economic conditions, we found *Politics, European Union,* and the *National Retirement System/INPS* (see their negative signs in Table 3).

It is unsurprising to note a significant negative Granger causality between the Covid keyword and the consumer evaluation of the economic climate. This implies that as the Covid term becomes more prevalent and widespread in online discussions, consumers' assessments and expectations of the Italian economic situation become increasingly pessimistic, with a bleak outlook on future employment prospects.

Another interesting result is the strong negative Granger causality between the keywords *educational degree, unemployment, purchase,* and the Climate's Future index, describing the expectations on the Italian economic situation, households' financial situation, unemployment expectations, and savings. The findings suggest that when education takes center stage in online discussions, consumers tend to have a less optimistic outlook on the future. This could indicate a decrease in trust regarding the effectiveness of education and obtaining degrees in shaping future prospects. Similarly, there seems to be a more pessimistic response from consumers when the news reports frequently and consistently about purchasing ability and fighting unemployment. This finding is consistent with another result that indicates a negative Granger causality between media coverage of the national retirement system and individuals' current and future personal situations. If we consider that the unemployment rate in Italy from 2016 to 2020 went from 11.7 to 9.3% (https://www.statista.com/statistics/531010/unemployment-rate-italy/, accessed April 16, 2021), this pessimistic view is consistent with the inception and spread of an economic downturn, partially ascribed to the Covid-19 pandemic. This result seems to confirm the negativity bias: when the present looks good, people feel more optimistic about the current and future economy[10].

In summary, the findings presented in Table 2 indicate that 27% of the selected keywords have a Granger-causal relationship with the aggregate Climate. This percentage is consistent with the results obtained when evaluating Granger causality for the Current dimension of the survey. These results suggest that a significant portion of the selected keywords can be used to predict changes in the Climate dimension, providing valuable insights for future research and decision-making. Our tests indicate that a higher number of keywords could impact how consumers perceive the Future situation. However, the most significant impact appears to be on the personal climate, as evidenced by 61% of significant Granger causality tests.

Finally, it is worth noting that the sentiment variable exhibits a significant correlation solely with the Personal component of the Consumer Confidence Index.

Table 3 provides a breakdown of the nine questions and offers a more granular view of the impact of ERKs on the adjusted consumer confidence index. Five of the nine questions pertain to the present circumstances, either of the household or the country, while the remaining four relate to the future expectations of both the nation and the consumers themselves. The importance of some keywords seems to be more impactful when associated with the single questions than with the aggregate climate measures presented in Table 2. In particular, all the keywords are strongly significant and improve the predictability of the household's economic situation index. Moreover, keywords related to Covid-19 appear highly predictive of consumers' current and future evaluation of the household economic situation and future unemployment.

In line with the findings presented in Table 2, it appears that ERKs have a greater influence on current assessments than on future projections. This is aligned with the current debate in the literature on consumer confidence, as it is still unclear whether surveys merely reflect current or past events or provide useful information about the future of household spending[8].

The Granger causality tests for sentiment indicate significance only for the second question, which pertains to the assessment of the household's economic situation.

**Forecasting.** As an additional step in our analysis, we conducted a forecasting exercise to examine the predictive capabilities of our new indicators in forecasting the Consumer Confidence Index. Our sample size is limited, which means that our analysis only serves as an indication of the potential of textual data to predict consumer confidence information. It is important to note that our findings should not be considered a final answer to the problem.

We performed monthly out-of-sample forecasting of the five main CCI indices (Climate Overall, Future Climate, Current Climate, Personal Climate, and Economic Climate).

| ERK series | Climate | | Personal | | Economic | | Current | | Future | |
|---|---|---|---|---|---|---|---|---|---|---|
| | Coef | Sign | Coef | Sign | Coef | Sign | Coef | Sign | Coef | Sign |
| Covid | 0.16 | − | 2.24 | − | 5.83*** | − | 0.36 | − | 0.26 | − |
| Lockdown | 0.09 | − | 2.23 | − | 0.22 | − | 0.38 | − | 0.05 | − |
| Unemployment | 2.54 | − | 5.86** | − | 1.41 | − | 1.02 | − | 3.05* | − |
| Quantitative easing | 2.65 | + | 3.70* | + | 1.36 | + | 1.12 | + | 2.58 | + |
| Economic crisis | 0.50 | − | 3.87* | − | 0.24 | − | 0.38 | − | 0.85 | − |
| Interest rate | 4.06** | − | 6.54** | + | 1.63 | + | 3.61* | + | 2.73* | + |
| Monetary politics | 3.48* | − | 6.04** | + | 1.92 | + | 2.99* | + | 3.23* | + |
| European community | 0.82 | − | 3.07* | − | 0.82 | − | 0.86 | + | 1.36 | − |
| Financial markets | 0.95 | + | 2.51 | + | 0.97 | + | 1.72 | + | 1.18 | + |
| EU | 1.68 | − | 10.33*** | − | 0.40 | − | 6.72** | −− | 0.91 | − |
| Unions | 0.08 | − | 1.93 | + | 0.81 | − | 0.57 | − | 0.12 | − |
| Strikes | 0.24 | − | 1.98 | + | 0.25 | + | 0.54 | + | 0.06 | + |
| Deficit | 0.73 | + | 5.16*** | + | 0.57 | + | 1.75 | + | 0.44 | + |
| Taxation | 0.23 | + | 2.02 | + | 1.21 | + | 0.48 | + | 0.49 | + |
| Btp-bot | 4.10** | + | 7.41*** | + | 2.55 | + | 6.08*** | + | 3.15* | + |
| Prices | 2.01 | − | 4.51** | + | 1.52 | − | 3.39* | + | 1.04 | + |
| Italian stock exchange | 3.80* | + | 6.28** | + | 1.82 | + | 6.19** | + | 2.46 | + |
| Bank of Italy | 2.08 | + | 4.41** | + | 0.72 | + | 0.88 | + | 1.78 | + |
| Smartworking | 0.55 | − | 2.29 | − | 0.84 | − | 0.84 | − | 0.21 | − |
| Junkbond | 2.05 | + | 6.04** | + | 1.41 | + | 1.07 | + | 2.66 | + |
| House | 2.66* | − | 1.94 | + | 2.86* | − | 0.71 | + | 2.64* | − |
| Purchase | 3.88* | − | 6.26** | + | 1.45 | − | 3.57* | + | 3.90** | − |
| Rent | 0.42 | − | 2.00 | − | 0.60 | − | 0.40 | − | 0.74 | − |
| Car | 1.92 | + | 1.94 | + | 3.22* | + | 1.21 | + | 1.24 | + |
| Incomes | 0.93 | + | 2.00 | + | 1.77 | + | 1.94 | + | 0.81 | + |
| Employee loans | 1.78 | + | 7.35*** | + | 1.02 | − | 1.37 | + | 2.25 | + |
| Loan | 2.00 | − | 6.74** | − | 1.51 | − | 4.57** | − | 1.97 | − |
| Family | 2.90* | − | 5.32** | + | 2.78* | − | 5.84** | + | 2.67 | + |
| Trust | 5.35** | + | 2.08 | − | 3.98** | + | 0.39 | + | 2.15 | + |
| Discomfort | 3.29* | + | 6.88*** | + | 1.32 | + | 3.88* | + | 2.76* | + |
| Job | 0.16 | + | 2.12 | + | 0.23 | + | 0.77 | + | 0.12 | + |
| PC | 1.68 | − | 2.09 | − | 0.42 | − | 1.76 | − | 0.13 | − |
| Politics | 1.55 | + | 6.90*** | − | 0.53 | + | 5.21** | + | 1.28 | + |
| Public sector | 3.18* | + | 7.97*** | + | 3.55* | + | 3.37* | + | 4.32** | + |
| Economy | 0.58 | − | 2.94* | − | 0.33 | − | 0.58 | − | 0.80 | − |
| Education degree | 6.64** | − | 5.65** | − | 3.64* | − | 1.30 | − | 7.45*** | − |
| Consumers | 0.81 | + | 3.72* | + | 0.88 | + | 2.10 | + | 1.27 | + |
| Needs | 1.13 | + | 3.31* | + | 2.54 | + | 1.34 | + | 1.85 | + |
| Global | 2.78* | + | 6.54** | − | 2.01 | + | 3.09* | + | 3.96** | + |
| Institutions | 0.35 | − | 2.01 | − | 1.61 | + | 0.47 | − | 1.28 | + |
| Retirement pension | 0.08 | + | 2.31 | + | 0.22 | + | 0.46 | + | 0.08 | + |
| Saving | 1.08 | + | 6.47** | + | 0.47 | + | 3.62* | + | 0.94 | + |
| Vacation | 0.10 | + | 2.00 | + | 0.23 | − | 0.70 | − | 0.06 | + |
| Competition | 3.34* | + | 8.58*** | + | 1.88 | + | 4.51** | + | 5.02** | + |
| Spread | 3.81* | + | 2.41 | + | 3.54* | + | 1.17 | + | 4.46** | + |
| Rating | 2.69 | + | 2.84* | + | 3.78** | + | 1.45 | + | 2.75* | + |
| Euro-group | 1.18 | − | 5.07** | + | 1.10 | − | 0.93 | + | 1.75 | − |
| Coronabond | 2.62 | − | 6.75** | + | 1.91 | − | 1.16 | + | 3.68* | − |
| Eurobond | 2.49 | − | 6.29** | + | 1.96 | − | 1.15 | + | 3.42* | − |
| Sure | 1.53 | − | 3.44* | + | 2.40 | − | 0.35 | + | 3.97** | − |
| European investment bank | 2.64 | + | 7.58*** | + | 1.15 | + | 1.34 | + | 3.10* | + |
| Oil | 0.99 | + | 2.24 | + | 0.93 | + | 0.88 | + | 0.51 | + |
| Gold | 3.90** | + | 2.10 | + | 4.85*** | + | 0.98 | + | 4.30** | + |
| Troika | 4.25** | − | 5.79** | + | 4.33** | − | 1.57 | + | 6.42** | − |
| Euro | 0.08 | + | 2.09 | + | 0.74 | + | 0.44 | + | 0.08 | + |
| Continued | | | | | | | | | | |

| ERK series | Climate | | Personal | | Economic | | Current | | Future | |
|---|---|---|---|---|---|---|---|---|---|---|
| | Coef | Sign | Coef | Sign | Coef | Sign | Coef | Sign | Coef | Sign |
| Italian public retirement system | 1.80 | – | 9.63*** | – | 0.84 | – | 3.38* | – | 2.41 | – |
| GDP | 0.47 | – | 2.55 | + | 0.45 | – | 0.51 | + | 0.37 | + |
| Confindustria (National Industrial Association) | 1.66 | – | 2.21 | – | 0.96 | – | 0.40 | – | 1.31 | – |
| Sentiment | 1.39 | + | 3.00* | + | 0.22 | + | 1.09 | + | 0.59 | + |

**Table 2.** Granger causality tests and cross-correlation signs of ERKs series and Consumer Climate (and its four components). $*p<.10$. $**p<.05$. $***p<.01$.

Considering that the Consumer Confidence surveys are administered during the first two weeks of the month, we chose as a benchmark model an autoregressive model with two lags AR(2):

$$y_{t+h} = \sum_{i=1}^{2} \phi_i y_{t+1-i} + \varepsilon_t, t = 1, \ldots, T$$

where $y_t$ is the target series, $h$ represents the number of steps ahead to forecast, $\phi_i$ is the $i$th coefficient of the autoregressive model of order $p=2$ and $\varepsilon_t$ represents a serially uncorrelated error white noise. We also tested other different models with more lags, without getting to better forecasting results.

To combine information from our large set of economic-related keywords and use it for forecasting, we created a factor-augmented autoregressive model (FAAR, also indicated as SBS ERK model) whose *h-step* ahead forecast is given by the following equation:

$$\widehat{y}_{T+h}^{FAAR} = \sum_{i=1}^{2} \widehat{\phi}_i y_{T+1-i} + \widehat{\xi}_j{}' \widehat{F}_T$$

where $F_t$ represents an $R \times 1$ vector of factors and $\xi$ a coefficient vector. All our models include the AR(2) component. Model estimation was carried out using an initial window of 60 weekly observations, expanded at each regression step. The total out-of-sample period comprised 30 monthly forecasts. The optimal number of factors was estimated dynamically through Partial Least Squares and using the Bayesian Information Criterion (BIC), with a maximum number of 4 factors[64].

We compared the forecasting results of our new models (which comprise the SBS scores of economic-related keywords) with those of our benchmark (the autoregressive model) and of two other models including the sentiment indicator (i.e., sentiment together with the AR(2) terms and the SBS ERK model also including sentiment).

Moreover, to go beyond the aggregate measures and get a complete picture of the SBS performance, we investigated the individual components—prevalence, diversity, and connectivity—separately.

Lastly, we considered a model based on BERT encodings[65] as an additional forecasting baseline.

In particular, this model was based on a neural network that processed encodings extracted by a pre-trained BERT model. In the following, the encodings extraction stage is first detailed, and then the neural network structure and its optimization are described.

Since the news articles considered in this work are written in Italian, we used a BERT tokenizer to preprocess the news articles and a BERT model to encode them; both pre-trained on a corpus including only Italian documents.

The BERT model for the computation of the encodings processes input vectors with a maximum of 512 tokens. Therefore, a strategy to handle vectors with more than 512 elements is necessary. In this work, we considered and compared two variants.

The first (referred to as BERT-truncated) considered only the first 30% of the tokens resulting from the tokenization procedure of the input news article. We truncated or padded the token vector with zeros to get 510 elements and added the classification [CLS] and separation [SEP] tags. The resulting vector was fed into a pre-trained BERT encoder, which computed a 768-element encoding vector for each token. Among these, we only considered the encoding of the [CLS] token to represent the news article, as it captures BERT's understanding at the news level.

In a second approach (referred to as BERT-chunk), we divided the token vector of each article into chunks of 510 elements, adding at the beginning and the end the [CLS] and [SEP] tags, respectively. The last chunk was padded to 512, if necessary. The BERT model then processed each chunk to extract the embeddings associated with the [CLS] tag, as in the BERT-truncated case. The embeddings of the [CLS] tags of all the chunks were then averaged to obtain a vector representing the full news article.

In both cases, the encodings of the [CLS] tokens for all the news articles in a week were averaged to obtain a vector summarizing the information for that week.

To nowcast CCI indexes, we trained a neural network that took the BERT encoding of the current week and the last available CCI index score (of the previous month) as input. The network comprised a hidden layer with ReLU activation, a dropout layer for regularization, and an output layer with linear activation that predicts the CCI index.

As with the other forecasting models, we implemented an expanding window approach to generate our predictions. Specifically, we started with an initial subset of data to train the neural network and make a first prediction for the next period. The training set window was subsequently expanded by including the next observation, and the process was repeated recursively.

| ERK series | Evaluation of the economic situation in Italy | Evaluation of the household economic situation | Evaluation of the household budget | Current opportunities for Savings | Current opportunities of purchasing durable goods | Expectations on the economic situation of Italy | Unemployment expectations | Expectations on the household economic situation | Future possibilities for savings |
|---|---|---|---|---|---|---|---|---|---|
| Covid | 0.52 | 7.16*** | 0.11 | 1.48 | 0.05 | 0.47 | 8.87*** | 6.41*** | 0.56 |
| Lockdown | 0.8 | 7.31*** | 0.19 | 1.53 | 0.01 | 0.07 | 5.51*** | 5.07*** | 0.44 |
| Unemployment | 0.75 | 6.57*** | 0.58 | 2.53 | 4.93** | 3.08* | 0.99 | 2.88* | 2.07 |
| Quantitative easing | 0.9 | 8.14*** | 0.12 | 1.74 | 1.04 | 2.81* | 0.23 | 0.4 | 0.91 |
| Economic crisis | 0.53 | 6.62*** | 0.13 | 5.15*** | 0.02 | 0.48 | 0.19 | 0.09 | 2.41 |
| Interest rate | 4.21** | 6.89*** | 0.14 | 3.63** | 2.44 | 0.82 | 0.54 | 0.4 | 6.2** |
| Monetary politics | 4.89*** | 8.84*** | 0.3 | 2.02 | 1.78 | 2.58 | 0.7 | 0.33 | 1.52 |
| European community | 1.05 | 6.57*** | 0.17 | 4.52*** | 2.89* | 1.08 | 0.39 | 0.26 | 1.02 |
| Financial markets | 1.02 | 7.97*** | 0.14 | 1.68 | 3.15* | 0.67 | 0.76 | 0.07 | 0.28 |
| EU | 2.42 | 5.93*** | 4.21** | 4.11** | 1.94 | 0.13 | 0.37 | 0.14 | 10.01*** |
| Unions | 1.41 | 6.96*** | 5.88** | 1.59 | 0.19 | 0.56 | 0.35 | 0.76 | 0.94 |
| Strikes | 1.21 | 7.24*** | 0.78 | 1.56 | 0.71 | 0.08 | 0.09 | 0.57 | 0.29 |
| Deficit | 0.85 | 6.2*** | 0.58 | 2.15 | 0.55 | 0.08 | 2.17 | 0.35 | 0.94 |
| Taxation | 1.93 | 5.55*** | 0.29 | 2.96* | 0.14 | 1.71 | 0.07 | 0.07 | 0.37 |
| Btp-bot | 3.36* | 5.94*** | 0.13 | 1.48 | 5.37** | 1.39 | 2.68 | 0.32 | 1.86 |
| Prices | 2.55 | 5.9*** | 1.47 | 3.35** | 6.32** | 0.67 | 1.4 | 0.07 | 0.3 |
| Italian stock exchange | 3.51* | 7.84*** | 0.26 | 3.5* | 2.3 | 0.92 | 0.74 | 0.41 | 2.4 |
| Bank of Italy | 0.54 | 7.09*** | 0.3 | 1.63 | 2.44 | 1.04 | 0.05 | 1.37 | 1.38 |
| smartworking | 2.39 | 6.89*** | 1.82 | 1.69 | 0.15 | 0.21 | 2.41 | 0.11 | 0.36 |
| Junkbond | 0.71 | 7.2*** | 0.52 | 1.72 | 3.08* | 3.59* | 0.26 | 0.88 | 1.27 |
| House | 0.61 | 7.11*** | 0.73 | 1.59 | 0.29 | 3.21** | 0.06 | 0.13 | 0.21 |
| Purchase | 2.33 | 7.2*** | 0.27 | 1.54 | 3.71* | 2.26 | 0.24 | 1.41 | 8.04*** |
| Rent | 3.18** | 7.03*** | 0.12 | 1.7 | 0.24 | 0.76 | 0.28 | 0.11 | 0.99 |
| Car | 7.59*** | 6.99*** | 2.16 | 3.46* | 0.02 | 0.91 | 0.96 | 0.63 | 1.08 |
| Incomes | 2.63 | 5.58*** | 0.4 | 2.01 | 3.06* | 0.53 | 0.19 | 0.95 | 0.73 |
| Employee loans | 0.59 | 7.04*** | 0.34 | 1.94 | 2.94* | 2.94* | 0.11 | 0.88 | 1.58 |
| Loan | 2.82* | 5.7*** | 1.01 | 1.62 | 1.82 | 0.59 | 0.27 | 0.25 | 3.73* |
| Family | 5.47** | 6.46*** | 3.06* | 1.87 | 1.58 | 0.74 | 2.57 | 2.22 | 6.56** |
| Trust | 4.26** | 7.4*** | 0.65 | 2.56 | 1.76 | 3.06* | 3.96** | 0.45 | 0.47 |
| Discomfort | 0.76 | 7.35*** | 0.22 | 3.8* | 2.1 | 2.27 | 0.11 | 0.5 | 1.14 |
| Job | 1.13 | 7.61*** | 0.21 | 2.58 | 0.02 | 0.08 | 0.58 | 0.1 | 1.42 |
| PC | 3.02* | 6.94*** | 4.03** | 1.48 | 0.06 | 0.92 | 1.39 | 0.14 | 1.65 |
| Politics | 1.58 | 8.6*** | 1.14 | 2.78* | 2.31 | 0.12 | 0.05 | 0.55 | 4.81** |
| Public sector | 2.58 | 6.9*** | 0.78 | 1.49 | 7.1*** | 6.17** | 0.74 | 0.46 | 1.99 |
| Economy | 0.58 | 8.47*** | 0.28 | 1.81 | 0.62 | 0.22 | 0.11 | 0.17 | 2.34 |
| Education/degree | 0.59 | 5.95*** | 0.33 | 5.92*** | 4.74** | 6.49*** | 2.35 | 2.29 | 1.14 |
| Consumers | 0.62 | 5.95*** | 1.56 | 1.53 | 1.48 | 1.36 | 0.25 | 0.75 | 0.2 |
| Needs | 2.08 | 6.41*** | 0.5 | 1.48 | 2.18 | 2.52 | 0.98 | 0.15 | 0.94 |
| Global | 0.53 | 5.72*** | 0.47 | 3.73* | 3.01* | 3.59* | 0.55 | 0.71 | 0.94 |
| Institutions | 0.81 | 6.92*** | 0.42 | 1.51 | 3.76* | 2.1 | 0.34 | 0.37 | 0.89 |
| Retirement pension | 0.61 | 5.78*** | 0.26 | 5.58*** | 0.33 | 0.14 | 0.15 | 0.38 | 0.18 |
| Saving | 0.54 | 5.95*** | 0.32 | 1.57 | 0.5 | 0.56 | 1.37 | 0.62 | 1.33 |
| Vacation | 3.34* | 5.59*** | 2.63 | 1.51 | 0.18 | 0.57 | 0.31 | 3.23* | 4.17** |
| Competition | 0.89 | 7.22*** | 1.02 | 4.56** | 2.28 | 1.78 | 1.69 | 1.8 | 4.4** |
| Spread | 2.32 | 6.3*** | 0.11 | 1.77 | 0.29 | 2.78* | 6.08** | 0.49 | 0.23 |
| Rating | 3.32** | 6.02*** | 0.27 | 5.08*** | 0.01 | 1.4 | 4.46** | 1.24 | 0.22 |
| Euro-group | 0.52 | 5.63*** | 0.19 | 1.48 | 1.73 | 2.2 | 0.74 | 1.23 | 0.37 |
| Coronabond | 0.68 | 6.96*** | 0.46 | 1.67 | 3.51* | 5.02** | 0.34 | 1.24 | 1.8 |
| Eurobond | 0.66 | 6.63*** | 0.55 | 1.63 | 3.06* | 5.01** | 0.37 | 1.4 | 1.61 |
| Continued | | | | | | | | | |

| ERK series | Evaluation of the economic situation in Italy | Evaluation of the household economic situation | Evaluation of the household budget | Current opportunities for Savings | Current opportunities of purchasing durable goods | Expectations on the economic situation of Italy | Unemployment expectations | Expectations on the household economic situation | Future possibilities for savings |
|---|---|---|---|---|---|---|---|---|---|
| Sure | 0.69 | 6.57*** | 1.68 | 1.52 | 1.53 | 7.14*** | 0.12 | 1.44 | 0.91 |
| European investment bank | 0.53 | 7.19*** | 1.1 | 2.71 | 1.73 | 2.87* | 0.78 | 2.7 | 0.53 |
| Oil | 0.67 | 5.9*** | 3.16* | 1.62 | 1.78 | 0.5 | 1.23 | 0.36 | 0.7 |
| Gold | 4.27** | 5.85*** | 2.74* | 3.69* | 6.78** | 4.36** | 2.22 | 0.53 | 0.19 |
| Troika | 1.39 | 6.99*** | 1.35 | 1.54 | 4.64** | 8.02*** | 2 | 2.51 | 1.03 |
| Euro | 0.64 | 6.53*** | 0.42 | 3.07* | 0.12 | 0.83 | 0.23 | 0.09 | 0.29 |
| National retire-ment system/INPS | 0.58 | 6.77*** | 0.6 | 1.84 | 2.69 | 2.13 | 0.06 | 0.49 | 5.46** |
| GDP | 0.79 | 6.78*** | 0.12 | 1.77 | 0.56 | 0.31 | 0.06 | 0.08 | 0.53 |
| Confindustria (National Industrial Asso-ciation) | 0.96 | 5.64*** | 1.01 | 3.23* | 0.28 | 1.07 | 0.07 | 0.27 | 0.19 |
| Sentiment | 0.91 | 6.15*** | 0.42 | 1.97 | 0.89 | 1.04 | 0.331 | 0.68 | 0.42 |

**Table 3.** Granger causality tests between ERKs series and single survey questions. $^{*}p<.10$. $^{**}p<.05$. $^{***}p<.01$.

Additionally, we tested a neural network architecture with recurrent layers to explicitly model temporal dependencies. However, the performance we obtained was worse than the non-recurrent version we reported in the result section. This is probably due to the limited number of training samples, which are insufficient to optimize the more complex recurrent model.

Table 4 illustrates the mean square forecasting errors (MSFEs) relative to the AR(2) forecasts. The numbers in the table represent the forecasting error of each model with respect to the AR(2) forecasting error. We used the Diebold-Mariano test[66] to determine if the forecasting errors of each model were statistically worse (in italic) than the best model, whose RMSFEs are highlighted in bold.

The empirical findings indicate that SBS ERK models produce the most accurate forecasts for Climate Overall, Personal, and Economic Climate, while adding sentiment leads to the best forecasting of Future Climate.

Looking at SBS components, we can notice that all of them are equally accurate in forecasting Personal Climate, while connectivity is the best performer also for Economic and Current Climate, for this second variable together with diversity. Notice that both AR and BERT models are always statistically different with respect to the best performer, while AR(2) + Sentiment performs worse than the best model for 3 variables out of 5.

The worse performance of the BERT models can be attributed to the insufficient number of training samples, which hinders the neural network's ability to learn the forecasting task and generalize to unseen samples. A much larger dataset would be required to effectively leverage the high dimensionality of BERT encodings and model the complex dependencies between news and CCI indexes. Interestingly, the BERT-chunk model performed approximately the same as the BERT-truncated one. This is in line with the idea that most of the relevant information of a news article is contained at its beginning or that online readers focus mainly on the headline and the lead[67].

| | Climate overall | Personal climate | Economic climate | Current climate | Future climate |
|---|---|---|---|---|---|
| AR(2) | *1.000**** | *1.000**** | *1.000**** | *1.000**** | *1.000*** |
| AR(2) + sentiment | *0.434** | *0.484*** | 0.455 | *0.471** | 0.426 |
| AR(2) + SBS ERK | **0.391** | **0.394** | **0.396** | 0.429 | 0.432 |
| AR(2) + SBS ERK + sentiment | *0.432** | *0.489*** | 0.415 | 0.439 | **0.418** |
| AR(2) + prevalence | *0.410*** | **0.394** | 0.407 | 0.421 | 0.432 |
| AR(2) + diversity | 0.422 | **0.394** | 0.402 | **0.403** | 0.463 |
| AR(2) + connectivity | 0.394 | **0.394** | **0.396** | **0.403** | 0.432 |
| BERT truncated | *1.076**** | *1.107**** | *0.992*** | *1.134**** | *1.089*** |
| BERT chunk | *1.104**** | *1.131**** | *1.060*** | *1.100**** | *1.141*** |

**Table 4.** Forecasting results. Bold figures indicate the best forecasting models. Italic figures are those for which the Diebold-Mariano test rejects the null hypothesis of equal predictive accuracy with respect to the best forecasting model at a significance level of 0.1 (*), 0.05 (**), or 0.01 (***).

## Discussion and conclusions

Our research sheds light on the importance of incorporating diverse data sources in economic analysis and highlights the potential of text mining in providing valuable insights into consumer behavior and market trends. Through the use of semantic network analysis of online news, we conducted an investigation into consumer confidence. Our findings revealed that media communication significantly impacts consumers' perceptions of the state of the economy. Figure 4 shows the economic-related keywords that can have a major role in influencing consumer confidence (those with the most significant Granger-causality scores, as presented in Section "Results").

Through a granular analysis of the dimensions of consumer confidence, we found that the extent to which the news impacts consumers' economic perception changes if we consider people's current versus prospective judgments. Our forecasting results demonstrate that the SBS indicator predicts most consumer perception categories more than the language sentiment expressed in the articles. ERKs seem to impact more the Personal climate, i.e., consumers' perception of their current ability to save, purchase durable assets, and feel economically stable. In addition, we find a disconnect between the ERKs' impact on the current and future assessments of the economy, which is aligned with other studies[68,69]. While the Consumer Confidence Index has often been considered a suitable predictor of economic growth and a good indicator of consumers' optimism about the current economy, short-term estimations may show deviations from long-term trends, likely caused by nonsystematic shocks.

Lastly, keywords associated with national or European political decisions seem to lead to more uncertainty and pessimism. This is consistent with other empirical evidence demonstrating how the conduct of politics—in our case, both at a national and European level—plays a role in determining how consumers feel about the economy's future in both the long and short run[9]. The higher prominence and predictive power of political keywords, both as it refers to economic and non-economic concerns, have been considered in past research among the key determinants of consumers' perception of the future of the economy[26,70].

These results are aligned with previous studies showing how exposure to uncertain information makes people feel uncertain and more pessimistic about their future[11,14,71]. People's reaction was more positive when keywords were associated with clear financial concepts (e.g., gold or monetary policy). When keywords were related to political discussions or concepts like rent, the role of Europe, or retirement, people's reaction was more negative. Interestingly, the keyword "gold" had an impact on determining consumer confidence in six of the nine questions: Evaluation of the Economic situation in Italy; Evaluation of the household economic situation; Evaluation of the household budget; Current Opportunities for Savings; Current Opportunities of Purchasing Durable Goods and Expectations on the economic situation of Italy. During economic downturns, such as the 2018 and 2019 recession in Italy, financial institutions often increase their holdings of gold as reserve assets. This may be due to the perception that gold is a safe and stable investment during times of economic uncertainty. As a result, consumers may view this move positively, as it signals financial stability and security within the institution. As demonstrated by a study commissioned by the IMF[72], macroeconomic announcements have a significant impact on both the price of gold and consumer confidence.

Even if exploratory in nature, our study suggests that news has important implications on consumer confidence during economic recessions, not only during an economic expansion, as suggested by recent research[37]. Overall, our models confirm the important role played by the media in shaping current judgments and future expectations[11], and the impact that national and European politics have on shaping these assessments[9].

This article investigates the antecedents of consumer confidence by analyzing the importance of economic-related keywords as reported on online news. After mining online Italian news over a period of four years, we found that most of the selected keywords impact how consumers perceive their personal economic situation.

Overall, this study offers valuable insights into the potential of semantic network analysis in economic research and underscores the need for a multidimensional approach to economic analysis. This study contributes to consumer confidence and news literature by illustrating the benefits of adopting a big data approach to describe current economic conditions and better predict a household's future economic activity. The methodology in this article uses a new indicator of semantic importance applied to economic-related keywords, which promises to offer a complementary approach to estimating consumer confidence, lessening the limitations of traditional survey-based methods. The potential benefits of utilizing text mining of online news for market prediction are undeniable, and further research and development in this area will undoubtedly yield exciting results. For example, future studies could consider exploring other characteristics of news and textual variables connected to psychological aspects of natural language use[73] or consider measures such as language concreteness[74].

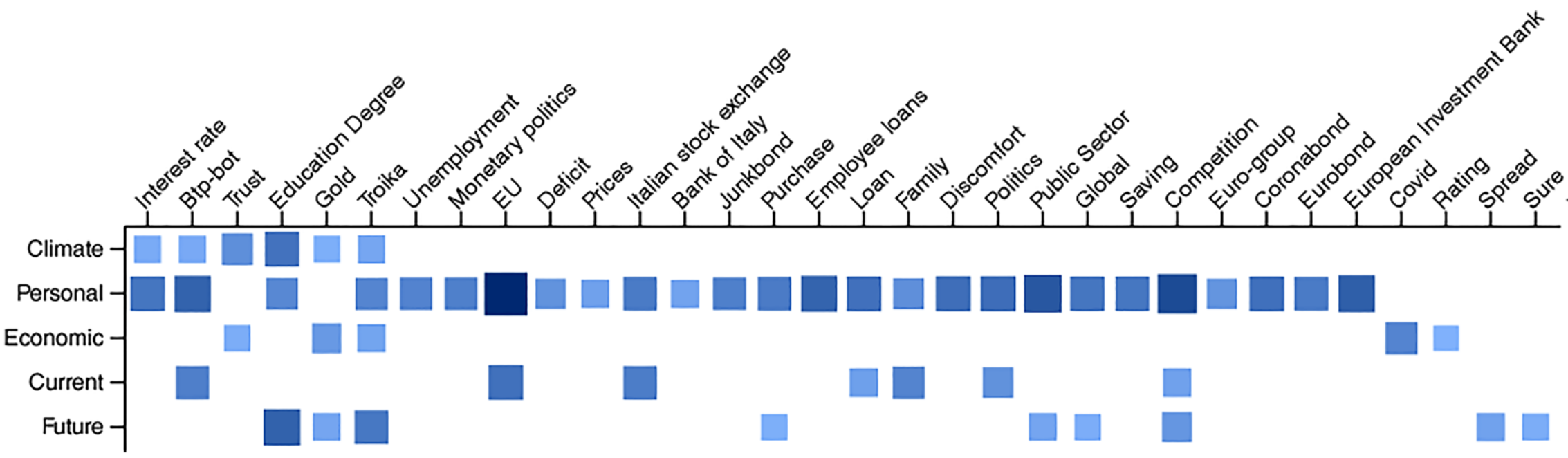

**Figure 4.** Most significant Granger causality results.

Finally, our research highlights the importance of media communication in shaping public opinion and influencing consumer behavior. As such, it is crucial for businesses and policymakers to be aware of the potential impact of media on consumer confidence and take appropriate measures to mitigate any negative effects.

## Data availability

The data that support the findings of this study are available from the author Barbara Guardabascio upon reasonable request. This refers to the numerical data resulting from the analysis of the news articles and the trained BERT models. However, the authors are not allowed to share the raw news data provided by Telpress International B.V. These data are the property of the company, and the authors have deleted them after the analysis.

## Acknowledgements
The authors wish to thank Vincenzo D'Innella Capano, CEO of Telpress International B.V., and to Lamberto Celommi, for making the news data available. The computing resources and the related technical support used for this work were provided by CRESCO/ENEAGRID High Performance Computing infrastructure and its staff. CRESCO/ENEAGRID High Performance Computing infrastructure is funded by ENEA, the Italian National Agency for New Technologies, Energy and Sustainable Economic Development and by Italian and European research programs.

## Author contributions
A.F.C.: Conceptualization; Methodology; Software; Formal Analysis; Data Curation; Writing—Original Draft; Writing—Review & Editing. F.G.: Conceptualization; Writing—Original Draft; Writing—Review & Editing.

B.G.: Conceptualization; Methodology; Software; Formal Analysis; Data Curation; Writing—Original Draft; Writing—Review & Editing. G.C.: Methodology; Software; Writing—Review & Editing. F.R.: Conceptualization; Funding acquisition; Writing—Review & Editing.

## Competing interests

The authors declare no competing interests.

## Additional information

**Correspondence** and requests for materials should be addressed to A.F.C.

**Reprints and permissions information** is available at www.nature.com/reprints.

**Publisher's note** Springer Nature remains neutral with regard to jurisdictional claims in published maps and institutional affiliations.